\documentclass{article}
\usepackage{breakurl}             
\usepackage{amsmath}
\usepackage{amsfonts}
\usepackage{amsthm}
\usepackage{amssymb}
\usepackage{algorithm}
\usepackage{algorithmicx}
\usepackage{algpseudocode}
\usepackage{tikz}
\usetikzlibrary{automata,positioning}
\usepackage[english]{babel}
\newtheorem{theorem}{Theorem}

\newtheorem{lemma}[theorem]{Lemma}
\newtheorem{definition}[theorem]{Definition}
\newtheorem{example}[theorem]{Example}

\usepackage{lineno,hyperref}
\modulolinenumbers[5]

\begin{document}


\title{Parameterized Verification of Coverability in Infinite State Broadcast Networks}

\author{A.R. Balasubramanian \\ Technical University of Munich \footnote{Work done when the author was a student at Chennai Mathematical Institute, India. A preliminary version of this paper appeared in GandALF 2018.} \\ ayikudir@in.tum.de}
\date{}
\maketitle



\begin{abstract}
Parameterized verification of coverability in broadcast networks with finite state processes has been studied for different types of models and topologies. In this paper, we attempt to develop a theory of broadcast networks in which the processes can be well-structured transition systems. The resulting formalism is called \textit{well-structured broadcast networks}. For various types of communication topologies, we prove the decidability of coverability in the static case, i.e, when the network topology is not allowed to change. We do this by showing that for these types of static communication topologies, the broadcast network itself is a \textit{well-structured transition system}, hence proving the decidability of coverability in the broadcast network. We also give an algorithm to decide coverability of well-structured broadcast networks when reconfiguration of links between nodes is allowed. Finally, with minor modifications of this algorithm we prove decidability of coverability when the underlying process is a pushdown automaton. 

\end{abstract}



\section{Introduction}

Specification and verification of infinite-state systems is a challenging task. Over the last two decades, various techniques have been proposed for checking safety and other properties of such systems, with one of the most prominent among them being the concept of a \textit{well-structured transition system} \cite{wsts,wqts}. A well-structured transition system is a transition system equipped with a well-quasi ordering on its states. Under some mild assumptions on the transition system, it is known that coverability in such systems is decidable.

Parameterized verification comprises of studying networks formed of anonymous agents executing the same code which interact with each other through some medium of communication, like broadcast, rendez-vous and shared variables \cite{crowd,reason,stochasticreg}. Its aim is to certify the correctness of all instances of the model, independently of the (parameterized) number of agents. Such problems are usually phrased in terms of infinite-state systems, to which technqiues from infinite-state verification theory can be applied. Indeed, a lot of results on parameterized verification prove that the underlying infinite state space of networks is a well-structured transition system. \cite{static, clique,register,time}

Broadcast networks are a formalism introduced in \cite{static}, in which the agents can broadcast messages simultaneously to all its neighbors. The number of agents and the communication topology are fixed before the start of the execution. Parameterized verification of such systems involves checking whether a specification holds irrespective of the number of agents or the communication topology. One of the prominent specifications considered in literature for such systems is the problem of coverability: \textit{does there exist an initial configuration from which at least one agent may reach a particular state}. In \cite{static}, the authors prove that the coverability problem for broadcast networks is undecidable even when the agents are finite state processes. Also, undecidability  has been proven for broadcast networks restricted to bounded-diameter topologies \cite{clique} and decidability has been proven for bounded-path topologies \cite{static},
bounded-diameter and degree topologies \cite{clique}, and clique topologies. Further, when we allow \textit{reconfigurations} of links in the underlying communication topology, there exists a polynomial time algorithm to decide coverability of broadcast networks comprising of finite-state processes \cite{mobile}.
This result perhaps seems surprising, since the reconfigurable case looks like a generalization of the static case. There has also been some work in extending the results of parameterized verification from the finite-state case to probabilistic automata \cite{prob,probtime} and timed automata \cite{time}. With the theory of broadcast networks having been explored for these various types of models, it seems natural to try to develop a theory of broadcast networks with well-structured transition systems as the underlying processes.

In this paper we study the coverability problem for broadcast networks where each process can be a \textit{labelled well-structured transition system}. In such systems, the underlying process itself can have infinite states. We call such systems \textit{well-structured broadcast networks}. We prove that the coverability problem is decidable for various classes of restricted topologies in this setting. In particular, we prove decidability for the set of all clique topologies, the set of all path-bounded topologies and the set of all topologies with bounded diameter and degree. We show that for these sets of topologies with well-structured transition systems as processes, the underlying state space of networks is itself a well-structured transition system. We also give an algorithm for deciding the coverability of a configuration for well-structured broadcast networks when reconfiguration of edges is permitted between the interacting agents.
This algorithm can also be modified slightly to yield an algorithm for coverability when the underlying process is a \textit{pushdown automaton}.\\

\textbf{Acknowledgements: } I am extremely grateful to Nathalie Bertrand and Nicolas Markey for useful discussions on the topic and also for assisting in the preparation of this paper. I would like to thank Igor Walukiewicz and B. Srivathsan for their help in arranging the necessary funding. I would also like to thank Thejaswini K.S and Mirza Ahad Baig for comments on early drafts of this paper and the anonymous reviewers for their valuable feedback, which greatly improved the presentation
of the paper.

\section{Well-structured broadcast networks}

In this section, we recall results about \textit{well-structured transition systems} \cite{wsts,wqts} and use them to 
define \textit{well-structured broadcast networks}. We also introduce the \textit{reconfiguration} semantics for such networks as a way
of modelling link changes that might occur in the underlying communication topology.

\subsection{Well-structured transition systems}

\begin{definition}
	A well-quasi ordering (wqo) $\le$ on a set $X$ is a reflexive, transitive binary relation such that any infinite sequence of elements $x_0,x_1,\cdots$ contains an increasing pair $x_i \le x_j$ with $i < j$.
\end{definition}

\begin{definition}
	A labelled well-structured transition system (labelled WSTS) is a tuple $(S,\Sigma,S_0,R,\le)$ where 
	
	\begin{itemize}
		\item $S$ is a set of configurations
		\item $\Sigma$ is a finite set of symbols called the alphabet
		\item $R \subseteq S \times \Sigma \times S$ is the transition relation
		\item $S_0$ is the set of initial configurations
		\item $\le \ \subseteq S \times S$ is a well-quasi order between states such that: 
		\begin{itemize}
			\item $\le$ is compatible with $R$, i.e., if $s_1 \le t_1$ and $(s_1,a,s_2) \in R$, 
			then there exists $t_2$ such that $s_2 \le t_2$ and $(t_1,a,t_2) \in R$
		\end{itemize}
	\end{itemize}
	
\end{definition}

To simplify notation, sometimes we will write $s \xrightarrow{a} s'$ to denote that $(s,a,s') \in  R$. Further we will say that a transition labelled by $a$ is enabled at a configuration $s$ iff there exists $s'$ such that $(s,a,s') \in R$.

Note that our definition of labelled WSTS is robust in the sense that if we restrict the WSTS to transitions of a particular label, we still get a WSTS. A WSTS is called \textit{finitely branching} if 
for each $s \in S$, there are only finitely many
transitions of the form $(s,a,s') \in R$. For simplicity of proofs, we will restrict ourselves to only finitely branching WSTS in this paper.

We call a set of configurations $I \subseteq S$, \textit{upward-closed} if $x \in I$ and $y \ge x$ implies $y \in I$. To any subset $I \subseteq S$,
we define $\uparrow I = \{x: \exists y \in I, \;
x \ge y\}$. In particular a set $I$ is upward-closed iff $I = \uparrow I$. 
A \textit{basis} for an upward-closed set $I$, is a set $I^b$ such that $I = \uparrow I^b$. It is known that for a wqo, every upward-closed set has a finite basis.

Given a set of configurations $I$, denote by $\mathit{pre}(I)$ the set $\{s' \in S: (s',a,s) \in R, \text{ for some } a \in \Sigma, s \in I\}$. For $i > 0$, let $\mathit{pre}^i(I):= \{s' \in S: (s',a,s) \in R, \text{ for some } a \in \Sigma, s \in \mathit{pre}^{i-1}(I)\}$ and let $\mathit{pre}^*(I):= \bigcup_{i \in \mathbb{N}} \; \mathit{pre}^i(I)$. Note that by our definition of a WSTS, if $I$ is upward closed then $\mathit{pre}(I)$ is upward closed as well. We will sometimes write $s \rightarrow s'$ to mean that $s \in \mathit{pre}(s')$ and $s \xrightarrow{*} s'$ to mean that $s \in \mathit{pre}^*(s')$. A labelled WSTS is said to have \textit{effective pre-basis} if given a finite basis for the upward-closed set $I$, we can compute a finite basis for the set $\mathit{pre}(I)$.

The \textit{coverability} problem for labelled WSTS is the following: Given a configuration $s$, decide if there exists  $s'$ and $s_0$ such that $s_0 \in S_0$, $s' \ge s$ and $s_0 \xrightarrow{*} s'$.

From \cite{wqts,wsts} it is known that 

\begin{theorem}
	Coverability is decidable for labelled WSTS with effective pre-basis and a decidable wqo.
\end{theorem}

The idea behind the proof is as follows: Given a configuration $s$, we compute the following sequence of upward-closed sets: $U_0 = \uparrow s$ and $U_{i+1} = \mathit{pre}(U_i)$. This sequence will eventually saturate to some $U_m$ which will give us a finite basis for $\mathit{pre}^*(U_0)$. Checking whether $s$ can be covered now amounts to checking if there is at least one initial configuration in $\uparrow U_m$. 

Common examples of labelled WSTS include: Any finite state system, vector addition systems with states (VASS), Petri nets with reset arcs, Petri nets with transfer arcs and lossy counter machines.

A labelled WSTS might be an infinite state system and so it is infeasible to describe the entire set of configurations in an explicit way.
Usually, a labelled WSTS $(S,\Sigma,S_0,R,\le)$ is given by means of a finite description $(Q,\Sigma,Q_0,\Delta,\cdots)$. The 
finite description may have additional structure like counters, causal relations etc. The structure of the relation $\Delta$ depends on the type of labelled WSTS that it describes.

\begin{example}
	Let $(Q,\Sigma,Q_0,\Delta,V)$ be a vector addition system with states (VASS) where $Q$ is a finite set of states, $\Sigma$ is a finite alphabet, $Q_0$ is a set of initial states, $V$ is a finite set of vectors over $\mathbb{Z}^d$ 
	(for some $d$) and $\Delta$ is of the form $\Delta \subseteq Q \times \Sigma \times V \times Q$. This describes a labelled WSTS $(S,\Sigma,S_0,R,\le)$ where
	$S$ is the set of all configurations, i.e., $S = \{(p,u): (p,u) \in Q \times \mathbb{N}^d\}$, $S_0 = \{(p,u): (p,u) \in Q_0 \times \{0\}^d\}$ and $\le$ is the usual product ordering on $\mathbb{N}^d$. The transition
	relation $R$ is defined in the following manner: 
	$((p,u),a,(q,w)) \in R$ iff $\exists  v \in V,
	(p,a,v,q) \in \Delta$ such that $u+v \ge 0$ and $w = u+v$. In this case we see that each transition $((p,u),a,(q,w)) \in R$ is 
	described by a tuple $(p,a,v,q) \in \Delta$.
\end{example}

For the rest of this paper we will assume that every labelled WSTS will be given by means of a finite description. Hence we assume that operations of the form: \emph{Given $a \in \Sigma$, choose a minimal configuration $c \in S$ such that a transition labelled by $a$ is enabled at $s$} (or) \emph{Given $a \in \Sigma$ and $c \in S$ choose a configuration $c'$ such that $(c,a,c') \in R$} (or) \emph{Delete all transitions which are not labelled by $a$}, are decidable by means of the given finite description $P$.

\begin{example}
	If $(Q,\Sigma,Q_0,\Delta,V)$ is a VASS which describes a labelled WSTS $(S,\Sigma,S_0,R,\le)$ it is clear that if $a \in \Sigma$ and $(p,a,v,q) \in \Delta$ such that $v = (v_1,\dots,v_d)$, then the configuration $c = (p,u)$ where $u = (\min(0,-v_1),\dots,$ $\min(0,-v_d))$ is a minimal configuration such that there is a transition labelled by $a$ enabled at $c$. Further, suppose for a letter $a \in \Sigma$ and a configuration $c = (p,u)$, we want to construct a configuration $c'$ such that $(c,a,c') \in R$. It is clear that this can be accomplished by selecting a tuple of the form $(p,a,v,q) \in \Delta$ with $u \ge (\min(0,-v_1),\dots,$ $\min(0,-v_d))$ and then setting $c' = (q,u+v)$.
\end{example}

\subsection{Well-structured broadcast networks}

In this section, we define \textit{well-structured broadcast networks} and also introduce the \textit{reconfiguration} semantics.

Throughout the paper, we fix a finite alphabet $\Sigma$.
Let the set of symbols $\{!!a: a \in \Sigma\}$ be 
denoted by $\Sigma_b$ and let the set of symbols
$\{??a: a \in \Sigma\}$ be denoted by $\Sigma_r$.

\begin{definition}
	A \textbf{process} is a labelled well-structured transition system $\mathcal{P} = (S,\Sigma_b \cup \Sigma_r,S_0,R,\le)$.
\end{definition}

A well-structured broadcast network consists of several copies of a single process $\mathcal{P}$. Each configuration of such a network is 
an undirected graph in which each node is labelled by a configuration $s \in S$. Intuitively, the labels $!!a$ and $??a$ correspond to
broadcasting and receiving messages according to the topology specified by the underlying graph. Formally,

\begin{definition}
	An $S$-graph is a graph $G = (V,E,L)$ where $L$ is a labelling function $L: V \to S$.
\end{definition}

An $S$-graph represents an undirected graph in which each node $v \in V$ is executing the same process $\mathcal{P}$ and is currently in the
configuration $L(v)$.

We now use the notion of a process to define a transition system called the \textit{well-structured broadcast network}.

\begin{definition}
	
	Given a process $\mathcal{P} = (S,\Sigma_b \cup \Sigma_r,S_0,R,\le)$, a \textit{well-structured broadcast network} is a tuple $\mathit{BN}(\mathcal{P}) = (\Theta,\Theta_0,\rightarrow)$, where 
	
	\begin{itemize}
		\item $\Theta$ is the set of all finite $S$-graphs
		\item $\Theta_0$ is the set of all finite $S_0$-graphs and 
		\item $\rightarrow$ is defined as follows: If $\theta = (V,E,L)$ and $\theta' = (V,E',L')$, then $\theta \xrightarrow{a} \theta'$ iff
		
		\textbf{Broadcast: } $E = E'$ and $\exists v \in V$ such that
		\begin{enumerate}
			\item $(L(v),!!a,L'(v)) \in R$
			\item $(L(u),??a,L'(u)) \in R$ for every node $u$ connected to $v$
			\item $L'(w) = L(w)$ for every other node $w$
		\end{enumerate}
		
	\end{itemize}
	
\end{definition}

If $\theta_0 \in \Theta_0$, then $\theta_0$ will be called an initial graph.
Whenever the process $\mathcal{P}$ is clear from the context, we refer to the broadcast network only by $\mathit{BN}$.

The well-structured broadcast network can be thought of as follows: 
We have a graph in which each vertex runs a copy of the process $\mathcal{P}$ and the current 
label of the vertex $v$ denotes the configuration of the process at $v$.
At each time step, a process in some vertex $v$ chooses to broadcast a message ($!!a$)
and it is received ($??a$) by all its neighbors $u$. 

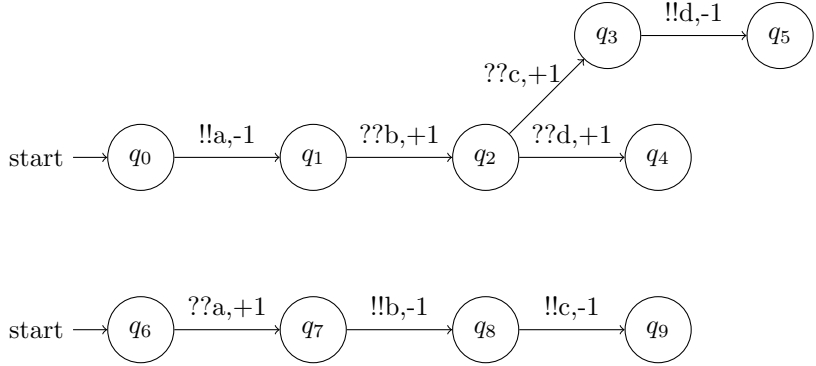
\begin{figure}
	\begin{center}
		\begin{tikzpicture}[->,node distance = 1.4cm]
		\node[state,initial] (q_0) {$q_0$};
		\node[state] (q_1) [right = of q_0] {$q_1$};
		\node[state] (q_2) [right = of q_1] {$q_2$};
		\node[state] (q_3) [above right = of q_2] {$q_3$}; 
		\node[state] (q_4) [right = of q_2] {$q_4$};
		\node[state] (q_5) [right = of q_3] {$q_5$};
		\node[state,initial] (q_6) [below = of q_0] {$q_6$};
		\node[state] (q_7) [right = of q_6] {$q_7$};
		\node[state] (q_8) [right = of q_7] {$q_8$};
		\node[state] (q_9) [right = of q_8] {$q_9$};
		
		\draw (q_0) edge[above] node{!!a,-1} (q_1);
		\draw (q_1) edge[above] node{??b,+1} (q_2);
		\draw (q_2) edge[above] node{??c,+1 $\hspace{15pt}$} (q_3);
		\draw (q_2) edge[above] node{??d,+1} (q_4);
		\draw (q_3) edge[above] node{!!d,-1} (q_5);
		\draw (q_6) edge[above] node{??a,+1} (q_7);
		\draw (q_7) edge[above] node{!!b,-1} (q_8);
		\draw (q_8) edge[above] node{!!c,-1} (q_9);
		\end{tikzpicture}
		\caption{Example of a process}
		\label{VASS process}
	\end{center}
\end{figure}

Figure \ref{VASS process} depicts a process whose specification is given by a VASS. The initial value of the counter is taken to be 1. If a transition for a receive symbol is not shown in the figure, it is assumed to go to a dead state.\\

Notice that this formulation of broadcast networks does not permit changes in links in the underlying topology.
To model such changes, we use the notion of \textit{reconfigurations}. A 
\textit{reconfigurable well-structured broadcast network} is a well-structured broadcast network in which along with \textit{broadcast} moves, we 
also allow transitions of the following kind: $\theta = (V,E,L) \rightarrow \theta' = (V,E',L')$ if \\

\textbf{Reconfiguration: } $L = L'$ and $E' \subseteq V \times V \setminus \{(v,v): v \in V\}$\\

Any reconfiguration corresponds to a non-deterministic change in the underlying network topology of the processes. We denote the resulting transition system by $\mathit{RBN}(\mathcal{P})$.

Given a well-structured broadcast network $\mathit{BN}$, the \textit{coverability} problem, given a configuration $s$, is to decide if there exists an initial graph such that by a series of transitions, we can reach a network topology in which at least one agent attains a configuration $s'$ which covers $s$. More formally, we consider the following problem: Given a configuration $s \in \mathcal{P}$, decide if there
exists $\theta \in \Theta$, $\theta_0 \in \Theta_0$ and $s' \ge s$ such that $\theta_0 \xrightarrow{*} \theta$ and $s'$ is the label of some process in $\theta$.
Notice that this is not the same as asking if $s$ is coverable in $\mathcal{P}$.

\begin{example}
	Consider a finite automaton with just two states $q,q'$ and a transition $q \xrightarrow{??a} q'$. Notice that the state $q'$ can never be reached in $\mathit{BN}(\mathcal{P})$. But when we treat this just as a labelled 
	transition system without the broadcast network semantics, it is clear that $q'$ can be reached from $q$ in the transition system $\mathcal{P}$. 
	To distinguish this, we refer to these two cases distinctly as coverability in $\mathit{BN}(\mathcal{P})$ and coverability in $\mathcal{P}$.
\end{example}

As a second comment we note that if a configuration is coverable in $\mathit{BN}(\mathcal{P})$, then it is also coverable in $\mathit{RBN}(\mathcal{P})$, but not vice versa. For example, consider the process $\mathcal{P}$ given in Figure \ref{VASS process}. We claim that in $\mathit{BN}(\mathcal{P})$, the state $q_4$ can never be covered. The reason is as follows: Suppose there is an execution in which some node $v$ reaches the state $q_4$. It is easy to see that the initial state of $v$ would have been $q_0$. To reach $q_4$ from $q_0$, the node $v$ should have received the message $d$ when it was as at state $q_2$ from some node $v'$. Therefore $v'$ should have been a neighbor of $v$ whose initial state was $q_0$. Since $v'$ had to transition from $q_0$ to $q_3$, it had to broadcast the message $a$ at some point. Consider the point in the run when $v'$ executed the transition $!!a$. Since $v$ was a neighbor of $v'$, $v$ had to receive the message $a$ sometime before it reached the state $q_4$. But upon receiving the message $a$, $v$ would have gone to a dead state, leading to a contradiction. Hence $q_4$ is not coverable in $\mathit{BN}$. However, we will see later that $q_4$ is indeed coverable when reconfigurations are allowed.

It is known that the coverability problem for well-structured broadcast networks $\mathit{BN}(\mathcal{P})$ is undecidable, even when $\mathcal{P}$ is a finite state transition system \cite{static}. As a way of overcoming undecidability, we will restrict the permissible set of underlying network topologies.

\section{Coverability problem for restricted topologies}

In this section, we investigate coverability in well-structured broadcast networks, where the set of all underlying graphs that we will consider will be restricted. In particular, we prove decidability results for three different classes of restricted topologies, namely bounded path topologies, clique topologies and bounded diameter and degree topologies. All these results could be seen as extensions of results that have been proved for 
finite state processes \cite{clique,static}.

As a first step, we define the \textit{induced subgraph ordering} between two configurations which will be used extensively to prove 
decidability in all three classes of topologies:

\begin{definition}
	Given two labelled graphs $\theta_1 = (V_1,E_1,L_1), \ \theta_2 = (V_2,E_2,L_2) \in \Theta$, define $\theta_1 \sqsubseteq \theta_2$ iff there exists an injection $h: V_1 \to 
	V_2$ such that $\forall u,v \in V_1$,
	
	\begin{itemize}
		\item $(u,v) \in E_1 \iff (h(u),h(v)) \in E_2$
		\item $L_1(u) \le L_2(h(u))$
	\end{itemize}
\end{definition}

In other words, the injection $h$ should preserve edges among vertices and also the order of their labels with respect to the well-quasi ordering. If such a $h$ exists then we will say that $h$ is an order preserving injection between $\theta_1$ and $\theta_2$.

\subsection{Bounded path topologies}

In this section, we prove that the coverability problem becomes decidable when we restrict to path bounded graphs. 
We will assume throughtout that a number $k$ is fixed.

In the sequel, given a labelled graph $\theta$, we will denote its vertex set by $V(\theta)$. Similarly, $E(\theta)$ and $L(\theta)$ will be used to denote the edge set and the label function of $\theta$ respectively.

\begin{definition} 
	A graph $G$ is called $k$-path bounded if the longest simple path in $G$ has length atmost $k$.
\end{definition}

Given a process $\mathcal{P}$, we can now define $k$-path bounded  broadcast networks by restricting the set of configurations in $\mathit{BN}(\mathcal{P})$ to $k$-bounded path topologies, i.e.,
we define a new transition system $\mathit{BN}^k(\mathcal{P}) = (\Theta^k,\Theta_0^k,\rightarrow)$, where $\Theta^k$ and $\Theta_0^k$ consists of only those configurations from
$\Theta$ and $\Theta_0$ which are $k$-path bounded. Notice that in this model, no reconfigurations are allowed between nodes.

	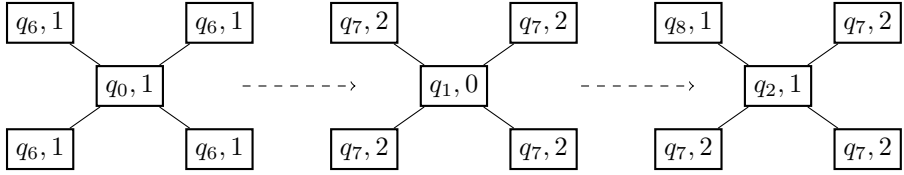
\begin{figure}
		\begin{tikzpicture}
		\begin{scope}[every node/.style={draw,rectangle,thick,minimum size = 5mm},node distance = 0.4cm]
		
		\node(n1) {$q_0,1$};
		\node(n2) [above left = of n1] {$q_6,1$};
		\node(n3) [above right = of n1] {$q_6,1$};
		\node(n4) [below left = of n1] {$q_6,1$};
		\node(n5) [below right = of n1] {$q_6,1$};
		\draw (n1) -- (n2);
		\draw (n1) -- (n3);
		\draw (n1) -- (n4);
		\draw (n1) -- (n5);
		
		\node(v1) [right = of n1, xshift = 3cm]{$q_1,0$};
		\node(v2) [above left = of v1] {$q_7,2$};
		\node(v3) [above right = of v1] {$q_7,2$};
		\node(v4) [below left = of v1] {$q_7,2$};
		\node(v5) [below right = of v1] {$q_7,2$};
		\draw (v1) -- (v2);
		\draw (v1) -- (v3);
		\draw (v1) -- (v4);
		\draw (v1) -- (v5);
		
		\node(w1) [right = of v1, xshift = 3cm]{$q_2,1$};
		\node(w2) [above left = of w1] {$q_8,1$};
		\node(w3) [above right = of w1] {$q_7,2$};
		\node(w4) [below left = of w1] {$q_7,2$};
		\node(w5) [below right = of w1] {$q_7,2$};
		\draw (w1) -- (w2);
		\draw (w1) -- (w3);
		\draw (w1) -- (w4);
		\draw (w1) -- (w5);

		\end{scope}
		
		\draw[dashed,->] (1.5,0) -- (3,0);
		\draw[dashed,->] (6,0) -- (7.5,0);
		
		\end{tikzpicture}
		\caption{Transitions between labelled graphs}
		\label{Fixed}
	\end{figure}

Figure \ref{Fixed} shows some transitions in the transition system $\mathit{BN}^2(\mathcal{P})$ for the process given in Figure \ref{VASS process}.\\

We will employ the theory of well-structured transition systems to prove that the coverability problem for $k$-path bounded broadcast networks is decidable. More specifically,
as a first step, we prove the following lemma.

\begin{lemma}
	The set of all $k$-path bounded configurations with the induced subgraph ordering is a well-quasi ordering.
\end{lemma}

\begin{proof}
	Follows from Ding's theorem (Theorem 2.2 in \cite{Ding}).
\end{proof}

As a next step, we prove that the induced subgraph ordering is \textit{compatible} with $\mathit{BN}^k(\mathcal{P})$.

\begin{lemma}
	For every $\theta_1,\theta_2,\theta_1' \in \Theta^k$ such that $\theta_1 \xrightarrow{a} \theta_2$ and $\theta_1 \sqsubseteq \theta_1'$, there exists $\theta_2'
	\in \Theta^k$ such that $\theta_1' \xrightarrow{a} \theta_2'$ and $\theta_2 \sqsubseteq \theta_2'$.
\end{lemma}

\begin{proof}
	Let $w$ be the vertex in $\theta_1$ which broadcasts the message $a$ and let $x_1,\cdots,x_p$ be the neighbors of $v$ which receive the message $a$. Let $h$ be an order preserving injection from $\theta_1$ to $\theta_1'$. Let $v = h(w)$ and $u_i = h(x_i)$ for each $i$. Since $\mathcal{P}$ is well-structured and since $L(\theta_1)(w) \le L(\theta_1')(v)$, it follows that there exists a transition $t'$
	labelled by $!!a$ which is enabled at $L(\theta_1')(v)$. Similarly,
	for each $i$, since $L(\theta_1)(x_i) \le L(\theta_1')(u_i)$, it follows that there
	exist transitions $t'_i$ labelled by $??a$ which are enabled at $L(\theta_1')(u_i)$ respectively. Since $h$ is an injection it follows that each $u_i$ is a neighbor of $v$. Hence, we can broadcast the message $a$ from the node $v$ and receive the message $a$ at the nodes $u_1,\cdots,u_p$ in the graph $\theta_1'$. Call the resulting graph $\theta_2'$. It is clear that the same injection $h$ is an order preserving injection between $\theta_1'$ and $\theta_2'$.
\end{proof}

As a final step, we prove the \textit{effective pre-basis} property.
Before proving so, we need some notations. Let $\mathcal{P} = (S,\Sigma_b \cup \Sigma_r, S_0,R,\le)$ be the given process and let $\mathit{BN}^k(\mathcal{P})$ be the associated $k$-path bounded broadcast network.
For a symbol $a \in \Sigma_b \cup \Sigma_r$ let $\mathcal{P}_a$ denote the transition system $\mathcal{P}$ restricted to only those transitions labelled by the symbol $a$. Notice that $\mathcal{P}_a$ is well-structured because $\mathcal{P}$ is well-structured. We assume that $\mathcal{P}_a$ inherits the effective-pre basis property from $\mathcal{P}$. For a set of configurations $C \subseteq S$, let $\mathit{pre}_{\mathcal{P}}(S)$ denote the set $\mathit{pre}(S)$ in the transition system $\mathcal{P}$ and $\mathit{pre}_{\mathcal{P}_a}(S)$ denote the set $\mathit{pre}(S)$ in the transition system $\mathcal{P}_a$.

As a first step to proving the effective pre-basis property we have the following lemma.

\begin{lemma} \label{firststep}
	Given a labelled graph $\theta \in \Theta^k$, we can effectively compute a finite basis for $\mathit{pre}(\uparrow \theta)$.
\end{lemma}

\begin{proof}
	Let $\theta = (V,E,L)$ and let $G = (V,E)$. For every node $v \in V$ and for every symbol $a \in \Sigma_b \cup \Sigma_r$, we assume that we can compute a finite basis for the set $\mathit{pre}_{\mathcal{P}_a}(\uparrow L(v))$, which we will denote by $B_a^v$. (We assume that this is accomplished by first deleting all transitions in $\mathcal{P}$ not labelled by $a$ and then using the effective basis property of $\mathcal{P}_a$ to compute a finite basis for $\mathit{pre}_{\mathcal{P}_a}(\uparrow L(v))$.
	
	Let $\mathbb{H} = \{H_1,H_2,\cdots,H_l\}$ be the set of all $k$-path bounded graphs such that each $H_i$ has one more vertex than $G$ and contains $G$ as an induced subgraph. Clearly $\mathbb{H}$ is finite.
	We will now describe two procedures whose outputs when taken together will constitute a basis for the set $\mathit{pre}(\uparrow \theta)$.\\

	The first procedure creates new \textit{labelled graphs} from the graph $G$ and is as follows:
	
	\begin{enumerate}
		\item Initialize a set $\mathtt{basis}_G$ to be empty.
		\item Choose a vertex $v \in V$ and a letter $a \in \Sigma$. Let $u_1,u_2,\cdots,u_p$ be the neighbors of $v$ in $G$.
		\item Choose a configuration $c_v$ from $B_{!!a}^v$.
		\item For each $u_j$, choose a configuration $c_{u_j}$ from $B_{??a}^{u_j}$.
		\item Construct the labelled graph $G_{\mathit{before}} = (G,L')$ as follows: $L'(v) = c_v, \ L'(u_j) = c_{u_j}$ for each $u_j$ and $L'(y) = L(y)$ if $y \notin \{v,u_1,\dots,u_p\}$.
		\item Choose a configuration $c_v'$ such that $(c_v,!!a,,c_v') \in R$.
		\item For each $u_j$, choose a configuration $c_{u_j}'$ such that $(c_{u_j},??a,c_{u_j}') \in R$.
		\item Construct the labelled graph $G_{\mathit{after}} = (G,L'')$ as follows: $L''(v) = c_v', \ L''(u_j) = c_{u_j}'$ for each $u_j$ and $L''(y) = L(y)$ if $y \notin \{v,u_1,\dots,u_p\}$.
		\item If $G_{\mathit{after}} \in \uparrow \theta$, add $G_{\mathit{before}}$ to the set $\mathtt{basis}_G$.
	\end{enumerate}
	
	Since the transition system $\mathcal{P}$ is assumed to be finitely branching, there are only finitely many choices to choose from in lines 6 and 7 of the above procedure. Further, as mentioned at the beginning of the paper, we assume that these choices can be effectively computed by means of the given finite specification for $\mathcal{P}$.
	
	It is clear by construction that $G_{\mathit{before}} \xrightarrow{a} G_{\mathit{after}}$ is a transition in $\mathit{BN}^k(\mathcal{P})$. Hence if $G_{\mathit{before}} \in \mathtt{basis}_G$ then $G_{\mathit{before}} \in \mathit{pre}(\uparrow \theta)$.\\

	The second procedure creates new \textit{labelled graphs} from the graphs in the set $\mathbb{H}$. Fix a graph $H \in \mathbb{H}$. From $H$, we construct labelled graphs in the following manner:
	
	\begin{enumerate}
		\item Initialize a set $\mathtt{basis}_{H}$ to be empty.
		\item Fix an injection $h: G \to H$ and a letter $a \in \Sigma$. Let $v$ be the vertex in $H$ which is not in the image of $G$, i.e., $v \notin h(G)$ and let $u_1,\cdots,u_p$ be the neighbors of $v$ in $H$.
		\item For the symbol $!!a$, choose a minimal configuration $c_v$ such that there is a transition labelled by $!!a$ enabled at $c_v$.
		\item For each $u_j$, choose a configuration $c_{u_j}$ from $B_{??a}^{u_j}$.
		\item Construct the labelled graph $H_{\mathit{before}} = (H,L')$ as follows: $L'(v) = c_v, \ L'(u_j) = c_{u_j}$ for each $u_j$ and
		$L'(y) = L(h^{-1}(y))$ if $y \notin \{v,u_1,\dots,u_p\}$.
		\item Choose a configuration $c_v'$ such that $(c_v,!!a,c_v') \in R$.
		\item For each $u_j$, choose a configuration $c_{u_j}'$ such that $(c_{u_j},??a,c_{u_j}') \in R$.
		\item Construct the labelled graph $H_{\mathit{after}} = (H,L'')$ as follows: $L''(v) = c_v', \ L''(u_j) = c_{u_j}'$ for each $u_j$ and 
		$L''(y) = L(h^{-1}(y))$ if $y \notin \{v,u_1,\dots,u_p\}$.
		\item If $H_{\mathit{after}} \in \uparrow \theta$, add $H_{\mathit{before}}$ to the set $\mathtt{basis}_H$.
	\end{enumerate}
	
	Once again it is clear by construction that $H_{\mathit{before}} \xrightarrow{a} H_{\mathit{after}}$ and so if $H_{\mathit{before}} \in \mathtt{basis}_{H}$ then $H_{\mathit{before}} \in \mathit{pre}(\uparrow \theta)$.\\

	Let $\mathtt{basis}_{\theta} := \mathtt{basis}_G \cup \left(\bigcup_{H \in \mathbb{H}} \mathtt{basis}_{H} \right)$. We claim that $\mathtt{basis}_{\theta}$ is a basis for the set $\mathit{pre}(\uparrow \theta)$. To this end, we show that if $\eta \in \mathit{pre}(\uparrow \theta)$ then there exists $\eta'$ such that $\eta' \sqsubseteq \eta$ and $\eta' \in \mathtt{basis}_{\theta}$. 
	
	Let $\eta := (V',E',L') \in \mathit{pre}(\uparrow \theta)$. Therefore, there should exist a transition from $\eta$ to some $\zeta := (V',E',L'') \in \uparrow \theta$. Let this transition be obtained by broadcasting $a$ from the vertex $w \in V'$ and receiving $a$ by all its neighbors $x_1,x_2,\cdots,x_l$. This means that there exist transitions $(L'(w),!!a,L''(w)) \in R$ and $(L'(x_j),??a,L''(x_j)) \in R$ for each $x_j$. Further $L'(y) = L''(y)$ for $y \notin \{w,x_1,\dots,x_l\}$.
	
	Recall that $\theta = (V,E,L)$ and $G = (V,E)$. Since $\zeta \in \uparrow \theta$ there exists an order preserving injection $h$ from $\theta$ to $\zeta$. We now have two cases:
	
	\begin{itemize}
		\item The node $w$ is in the image of $h$: Let $x_1,x_2,\cdots,x_p$ be the neighbors of $w$ in $\zeta$ which are in the image of $h$. 
		Let $v = h^{-1}(w)$ and $u_i = h^{-1}(x_i)$ for each $x_i$.
		
		Since $h$ is an order preserving injection and $h(v) = w$, we have $L(v) \le L''(w)$. By assumption, there exists a transition $(L'(w),!!a,L''(w)) \in R$.  Combining these two, we have $L'(w) \in \mathit{pre}_{\mathcal{P}_{!!a}}(\uparrow L(v))$. Since $B_{!!a}^v$ is a finite basis for $\mathit{pre}_{\mathcal{P}_{!!a}}(\uparrow L(v))$ it follows that there exists $c_v \in B_{!!a}^v$ and $c_v \le L'(w)$. Similar reasoning enables us to conclude that for each $u_i$, there exists $c_{u_i} \in B_{??a}^{u_i}$ and $c_{u_i} \le L'(x_i)$. 
		
		Since $c_v \in B_{!!a}^v$ it follows that there is a transition 
		$(c_v,!!a,c_v') \in R$ for some $c_v' \ge L(v)$. Similarly for each $u_i$, since $c_{u_i} \in B_{??a}^{u_i}$ it follows that there is a transition $(c_{u_i},??a,c_{u_i}') \in R$ for some $c_{u_i}' \ge L(u_i)$.
		
		Let $\eta' = (G,M')$ where $M'(v) := c_v, \ M'(u_i) = c_{u_i}$ and $M'(y) = L(y)$ for every $y \notin \{v,u_1,\dots,u_p\}$. Let $\zeta' = (G,M'')$ where $M''(v) = c_v', \ M''(u_i) = c_{u_i}'$ and
		$M''(y) = L(y)$ for every $y \notin \{v,u_1,\dots,u_p\}$.
		
		By construction it can be easily checked that $\eta' \xrightarrow{a} \zeta'$, $\zeta' \in \uparrow \theta$, $\eta' \in \mathtt{basis}_G$ and $h$ is an order preserving injection from $\eta'$ to $\eta$. Hence we have $\eta' \in \mathtt{basis}_G$ and $\eta' \sqsubseteq \eta$.

		\item The node $w$ is not in the image of $h$: Let 
		$x_1,\cdots,x_p$ be the neighbors of $w$ which are in the image of $h$. Let $u_i = h^{-1}(x_i)$ for each $x_i$. Consider the graph $H$ which is obtained from $G$ by adding one more vertex $v$ as a neighbor to $u_1,\dots,u_p$. Clearly $H \in \mathbb{H}$. 
		
		By assumption there exists a transition $(L'(w),!!a,L''(w)) \in R$. Let $c \le L'(w)$ be a minimal configuration such that there is a transition of the form $(c,!!a,c') \in R$ for some $c'$. Now, similar to the previous case we can obtain for each $u_i$, a configuration $c_{u_i} \in B_{??a}^{u_i}$ such that $c_{u_i} \le L'(x_i)$. For each $u_i$, since $c_{u_i} \in B_{??a}^{u_i}$ it follows that there is a transition $(c_{u_i},??a,c_{u_i}') \in R$ for some $c_{u_i}' \ge L(u_i)$.
		
		Let $\eta' = (H,M')$ where $M'(v) := c, \ M'(u_i) = c_{u_i}$ and $M'(y) = L(y)$ for every $y \notin \{v,u_1,\dots,u_p\}$. Let $\zeta' = (H,M'')$ where $M''(v) = c', \ M''(u_i) = c_{u_i}'$ and
		$M''(y) = L(y)$ for every $y \notin \{v,u_1,\dots,u_p\}$.
		Let $h'$ be the map $h'(w) = v$ and $h'(y) = h(y)$ for $y \neq w$.
		
		By construction it can be easily checked that $\eta' \xrightarrow{a} \zeta'$, $\zeta' \in \uparrow \theta$, $\eta' \in \mathtt{basis}_H$ and $h'$ is an order preserving injection from $\eta'$ to $\eta$. Hence we have $\eta' \in \mathtt{basis}_H$ and $\eta' \sqsubseteq \eta$. 
		
	\end{itemize}
\end{proof}

\begin{lemma} \label{effprebasis}
	If $C \subseteq \mathit{BN}^k(\mathcal{P})$ is an upward closed set and has a finite basis, then we can effectively compute a finite basis for $\mathit{pre}(C)$.
\end{lemma}

\begin{proof}
	Let $\mathbb{B} = \{\theta_1,\cdots,\theta_n\}$ be a finite basis of the upward-closed set $C$. By lemma \ref{firststep}, for each $\theta_i$, we can compute a set of labelled graphs $\mathtt{basis}_{\theta_i}$ such that $\mathtt{basis}_{\theta_i}$ is a basis for $\mathit{pre}(\uparrow \theta_i)$. It is then clear that the required basis for the set $\mathit{pre}(C)$ is simply $\bigcup_{1 \le i \le n} \ \mathtt{basis}_{\theta_i}$.
\end{proof}

\begin{theorem}
	Coverability in $k$-path bounded configurations is decidable.
\end{theorem}

\begin{proof}
	Let $s$ be the given configuration. Consider the graph $G$ with only one vertex whose label is $s$. It is clear that the configuration $s$ can be covered iff the graph $G$ can be covered in the transition system $\mathit{BN}^k$ under the induced subgraph ordering. But by the previous lemmas, we have shown that $\mathit{BN}^k$ is a \textit{well-structured transition system} under the induced subgraph ordering with an effective pre-basis.
	Therefore, coverability in $\mathit{BN}^k$ is decidable and this concludes the proof.
\end{proof}

Hence coverability in the broadcast semantics of $k$-path bounded topologies reduces to checking coverability in another WSTS!

\subsection{Clique topologies}

We prove a similar result for the set of all clique topologies. 
Let $\mathcal{P} = (S,\Sigma_b \cup \Sigma_r, S_0,R,\le)$ be the underlying process.

\begin{lemma} \label{cliquewqo}
	The set of all clique configurations forms a well-quasi ordering under the induced subgraph order.
\end{lemma}

\begin{proof}
	Recall that $S$ is the set of configurations of the process $\mathcal{P}$.
	We consider the poset $(M_f(S),\subseteq_{\le})$ where $M_f(S)$ is the set of all finite sub-multisets of $S$ and $\subseteq_{\le}$ is defined as
	\begin{equation*}
		S_1 \subseteq_{\le} S_2 \iff \exists \text{ an injection } h : S_1 \to S_2 \text{ such that } \forall s \in S_1, \ s \le h(s)
	\end{equation*}
	
	It is well known that if $(S,\le)$ is a wqo,
	then $(M_f(S),\subseteq_{\le})$ is also a wqo. Using this result we show that the set of all clique configurations are well-quasi ordered.

	Let $\theta$ and $\theta'$ be labelled clique configurations. Let $L_{\theta}$ be the \emph{multiset} $\{L(\theta)(v) : v \in V(\theta)\}$. Similarly let $L_{\theta'}$ be the multiset $\{L(\theta')(v) : v \in V(\theta')\}$. It is then clear that $L_{\theta} \subseteq_{\le} L_{\theta'}$ iff $\theta \sqsubseteq \theta'$. Since $\subseteq_{\le}$ is a well-quasi order, it follows that the set of all clique configurations forms a wqo under the induced subgraph ordering.
\end{proof}

The \textit{compatibility} property can be easily proved in an argument similar to the one given for $k$-path bounded graphs. The computation of pre-basis can be realized as follows: The algorithm given in the previous subsection, first selects a graph $\theta = (G,L)$ from the given basis $\mathbb{B}$ and then considers all $k$-path bounded graphs of size atmost $|G|+1$ which induce $G$ as a subgraph, after which it proceeds to construct a pre-basis from these $k$-path bounded graphs. We employ the same algorithm to construct a pre-basis for the clique topology as well, except in the first step, we replace the construction of the set of all $k$-path bounded graphs of size atmost $|G|+1$ which induce $G$, with the set of all clique graphs of size atmost $|G|+1$ which induce $G$. The proof of this algorithm follows from a similar proof given for the previous case. Hence we have,

\begin{theorem}
	Coverability in clique configurations is decidable.
\end{theorem}

\subsection{Graphs with bounded diameter and degree}

In this section we prove that the coverability problem is decidable when restricted to the space of all graphs with bounded diameter \emph{and}  bounded degree. First, we give a definition of diameter of a graph:

\begin{definition}
	The diameter of a graph $G$ is the maximum length of all shortest paths between any two vertices of $G$.
\end{definition}

It is known that the coverability problem for well-structured broadcast networks restricted to graphs of bounded diameter is undecidable, even when the underlying process is 
of finite state space \cite{clique}.
However we can regain decidability if along with bounded diameter, we also consider graphs of bounded degree.
To prove this, we use a non-trivial result of Hoffman and Singleton \cite{Moore}.
The result states for a fixed diameter $k$ and a degree $d$, the size of the largest (unlabelled) graph with diameter atmost $k$ and 
degree atmost $d$ is $M(k,d) = (k(k-1)^d-2)/(k-2)$. Hence, for finite state processes this immediately proves that the coverability problem is decidable.
But we can extend it in a straightforward way to the well-structured case as well. Recall that $S$ is the set of all configurations of the process $\mathcal{P}$. For a graph $G = (V,E)$ of bounded diameter and degree, 
consider the set $\mathit{Lab}(G) = \{G' \; | \; G' = (V,E,L), \ L: V \to S\}$, i.e., $\mathit{Lab}(G)$ is the set of all labelled graphs that can be obtained by
labelling the vertices in $G$ using labels from $S$.

\begin{lemma} \label{fixedwqo}
	For a fixed graph $G$ of diameter $k$ and degree $d$, the set $\mathit{Lab}(G)$ is a well-quasi ordering under the induced subgraph ordering.
\end{lemma}

\begin{proof}
	Suppose $G$ has $n$ vertices. Arbitrarily arrange the vertices in some order $v_1,\cdots,v_n$. Let $(S^n,\le^n)$ be the poset where $\le^n$ is defined as
	\begin{equation*}
		(s_1,\dots,s_n) \le^n (s_1',\dots,s_n') \iff s_1 \le s_1' \land s_2 \le s_2' \land \dots \land s_n \le s_n'
	\end{equation*}
	
	It is well know that if $(S,\le)$ is a wqo then $(S^n,\le^n)$ is also a wqo. Now notice then that if $\theta = (V,E,L)$ and $\theta' = (V,E,L')$ belong to $\mathit{Lab}(G)$ then 
	\begin{equation*}
		(L(v_1),\dots,L(v_n)) \le^n (L'(v_1),\dots,L'(v_n)) \implies \theta \sqsubseteq \theta'
	\end{equation*}
	
	and so the lemma immediately follows.
\end{proof}

For a fixed graph $G$, we can prove \textit{compatibility} and \textit{effective pre-basis} in a manner similar to the other cases.
Hence, we have

\begin{theorem}
	Coverability problem for $k$-bounded diameter and $d$-bounded degree graphs is decidable.
\end{theorem}

\begin{proof}
	Let $B = \{b_1,\cdots,b_m\}$ be a finite basis for the configuration space $S$ of the process $\mathcal{P}$ and let $s$ be the given configuration for which coverability needs to be determined.
	Let $G$ be a fixed graph on $n$ vertices and consider the set $$B_i = \{(b_{j_1},b_{j_2},\cdots,b_{j_{i-1}},s,b_{j_i},b_{j_{i+1}},\cdots,b_{j_n}): 
	\text{ each } b_{j_l} \in B\}$$ and let $B' = \bigcup_{1 \le i \le n} B_i$. Clearly the set $B'$ is finite. 
	
	The above properties imply that given a fixed graph $G$ of bounded diameter and degree, it can be decided if any configuration from $B'$ can be covered from any initial configuration in $\mathit{Lab}(G)$. But we know that the number of graphs with diameter $k$ and degree $d$ is finite. Hence, we can check if at least one configuration
	from $B'$ can be covered from any of these graphs and so the coverability problem is decidable for bounded diameter and degree graphs.
\end{proof}

\section{Coverability problem for reconfiguration semantics}

In the previous section, we proved decidability results for various classes of \textit{restricted static topologies}. However these systems permit no changes in the network topology of the system. In this section, we tackle the coverability problem for the reconfiguration semantics 
and prove that it is decidable. In particular, 
we present an algorithm which when given a process $\mathcal{P}$ and a configuration $s$, returns true iff the configuration $s$ can be covered in $\mathit{RBN}(\mathcal{P})$.

Let $\mathcal{P} = (S,\Sigma_b \cup \Sigma_r, S_0, R, \le)$ be the given process. 
Recall that $\Sigma$ is a fixed alphabet such that $\Sigma_b = \{!!a : a \in \Sigma\}$ and $\Sigma_r = \{??a : a \in \Sigma\}$. We assume that for every $a \in \Sigma$, we can compute a basis $C_a$ for the set of all configurations which have an enabled transition labelled by $!!a$. (This set is always finite, since the underlying order is a wqo). Notice that this computation concerns only the semantics of the transition system $\mathcal{P}$ and not that of $\mathit{RBN}$. For many systems such a computation will be fairly straightforward from the given finite specification of $\mathcal{P}$.

\begin{example}
	\begin{enumerate}
		\item If $\mathcal{P}$ is a finite state system, then for any letter $a$ we can search through the space of all configurations in 
		$\mathcal{P}$ and compute the transitions labelled by $!!a$.
		
		\item In a VASS, suppose $\{t^1,t^2,\dots,t^l\}$ are the set of all tuples labelled by $!!a$, i.e., each $t^i$ is of the form $(p^i,!!a,v^i,q^i)$. It is then clear that
		$C_a = \{(p^1,u^1),(p^2,u^2),\dots,(p^l,u^l)\}$ where each $$u^i = (\max(0,-v_1^i),\max(0,-v_2^i),\dots,\max(0,-v^i_n))$$
		
		\item Since transfer arcs and reset arcs in Petri nets have the same precondition as normal transitions, it follows that computation of 
		minimal configurations in these cases is similar to VASS \cite{resetarcs}.
	\end{enumerate}
\end{example}

A broadcast transition is a transition labelled by letters from $\Sigma_b$. Similarly, a receive transition is one labelled by letters from $\Sigma_r$. Further for each letter $a \in \Sigma$, we define $B_a$ to be the set of all broadcast transitions labelled by $!!a$. Similarly, 
we define $R_a$ to be the set of all receive transitions labelled by $??a$. Let $\mathit{Rec} = \bigcup_{a \in \Sigma} R_a$. 


\begin{algorithm}[h] 
	\caption{Coverability algorithm for reconfiguration semantics}
	\begin{algorithmic}[1]
		\State \textbf{Input: } A process $\mathcal{P} = (S,\Sigma_b \cup \Sigma_r, S_0, R, \le)$ and a config. $s \in \mathcal{P}$
		\State \textbf{Output: } Whether $s$ is coverable in the transition system $\mathit{RBN}(\mathcal{P})$\\
		
		\State $\mathcal{P}':= \mathcal{P} [R \leftarrow R \setminus \mathit{Rec}]$ 
		\Comment{{\footnotesize \color{blue} Remove all receive transitions to get $\mathcal{P}'$}}
		\State $\mathtt{SubAlp}:= \Sigma$\\
		\Repeat
		\State $\mathtt{AddT}:= \emptyset$ 
		\ForAll{$a \in \mathtt{SubAlp}$ } \Comment{{\footnotesize \color{blue} Look for symbols that can be broadcast in $\mathcal{P}'$}}
		\If{$\exists c \in C_a$ s.t. $c$ is coverable in $\mathcal{P}'$}
		\State $\mathtt{AddT}:= \mathtt{AddT} \cup R_a$ \Comment{{\footnotesize \color{blue} Store the receive transitions in $\mathtt{AddT}$}}
		\State $\mathtt{SubAlp}:= \mathtt{SubAlp} \setminus \{a\}$
		\EndIf
		\EndFor
		
		\State $\mathcal{P}':= \mathcal{P}'[R \leftarrow R \cup \mathtt{AddT}$] \Comment{{\footnotesize \color{blue} Add all transitions from $\mathtt{AddT}$ to $\mathcal{P}'$}}
		
		\Until{$\mathtt{AddT} = \emptyset$}\\
		
		\If{$s$ is coverable in $\mathit{\mathcal{P}'}$}
		\State \textbf{return true}
		\Else
		\State \textbf{return false}
		\EndIf
	\end{algorithmic}
\end{algorithm}

The coverability algorithm for $\mathit{RBN}$ is given in Algorithm 1. The algorithm proceeds as follows: As a first step, from the original process $\mathcal{P}$ we remove all transitions in $\mathit{Rec}$, to get a modified process $\mathcal{P}'$. (As mentioned in the beginning of the paper, we assume that this can be performed by appropriate operations on the finite specification of $\mathcal{P}$).
At each iteration of the main loop, for each letter $a$ we check if atleast one configuration from the set $C_a$ can be covered in the current process $\mathcal{P}'$. Intuitively, this means that some agent in the network can reach a configuration, from which it would be capable of broadcasting the letter $a$. At this point, we update the process $\mathcal{P}'$ by adding all the receive transitions labelled by $a$. (Once again made possible by appropriate operations on the finite specification). Whenever in the future, an agent wants to take a transition labelled by $??a$, it can do so now, because we can make another agent reach a configuration capable of broadcasting $a$, and then reconfigure the network, so that both these agents share an edge. (This is where the reconfiguration semantics of the network plays a prominent role in checking the coverability of a configuration). We keep doing this until no more transitions can be added, at which point we check if the required configuration is coverable in the resulting process obtained.

Notice that at any point in the algorithm, the transition system $\mathcal{P}'$ will always be a well-structured transition system. Indeed at the beginning of the code, $\mathcal{P}'$ is exactly the transition system obtained by removing all transitions labelled by symbols from $\Sigma_r$ from $\mathcal{P}$ and since $\mathcal{P}$ was a WSTS, $\mathcal{P}'$ will also remain a WSTS. Similarly, at each update of the $\mathcal{P}'$, we add all transitions of the form $??a$ for some symbol $a \in \Sigma$. Hence, the new transition system $\mathcal{P}'$ continues to be a WSTS. Further as mentioned above the operations in lines 4 and 15 are assumed to performed by means of the given finite specification.

The coverability tests in lines 10 and 18 refer to coverability in the transition system $\mathcal{P}'$. Also notice that whenever the algorithm increases the cardinality of the set $\mathtt{AddT}$, it decreases the size of $\mathtt{SubAlp}$ by 1. Since, the transitions added to $\mathtt{AddT}$ are labelled by symbols from $\mathtt{SubAlp}$ and since $\mathtt{SubAlp}$ is finite, it follows that eventually we can add no more transitions to $\mathtt{AddT}$. Therefore, line 16 of the algorithm will eventually become true and so the algorithm always terminates.

Let $\mathtt{AddT}_0 = \bigcup_{a \in \Sigma} B_a$ and for $i > 0$, let $\mathtt{AddT}_i$ be the contents of the set $\mathtt{AddT}$ at the end of the $i^{th}$ iteration of the outermost loop. Further, let $\mathcal{P}'_0 = \mathcal{P}[R \leftarrow R \setminus \mathit{Rec}]$ and for $i > 0$, let $\mathcal{P}'_i = \mathcal{P}'_{i-1}[R \leftarrow 
R \cup \mathtt{AddT}_i]$, i.e., $\mathcal{P}'_i$ denotes the process obtained at the end of the $i^{th}$ iteration of the outermost loop. Let the total number of iterations of the outermost loop be $w$. Hence we have a sequence of processes $\mathcal{P}'_0, \mathcal{P}'_1, \cdots, \mathcal{P}'_w$.


For a configuration $s$, we will say that $s$ is coverable in 
$\mathit{RBN}(\mathcal{P})$ if there exists an initial graph $\theta_0$ and a graph $\theta$ such that $\theta_0 \xrightarrow{*} \theta$ and
there exists a node $v \in \theta$ such that $L(\theta)(v) \ge s$.
The correctness of this algorithm follows by a series of lemmas.

\begin{lemma} \label{lemma1}
	If a configuration $s$ is reachable in $\mathcal{P}'_i$ for some $i$, then $s$ can be covered in the original reconfigurable broadcast network $\mathit{RBN}(\mathcal{P})$.
\end{lemma}

\begin{proof}
	
	Let $s$ be a configuration which is reachable in the transition system $\mathcal{P}'_i$. Further wlog, let $i$ be the first index such that $s$ is reachable in $\mathcal{P}'_i$. We will prove by induction on $i$ that the configuration $s$ is coverable in the broadcast network $\mathit{RBN}(\mathcal{P})$ as well.
	
	Suppose $i = 0$. Since $s$ is reachable in $\mathcal{P}'_0$, there exists a path $LP = s_0 \rightarrow s_1 \rightarrow \cdots \rightarrow
	s_{n-1} \rightarrow s_n = s$ in the transition system $\mathcal{P}'_0$. We prove the claim for $i = 0$ by a second induction on $n$. For the base case of $n = 0$, it is clear that $s_0$ is an initial configuration and so $s_0$ can be trivially covered in $\mathit{RBN}(\mathcal{P})$. Suppose $n > 0$. By our secondary induction hypothesis, the configuration $s_{n-1}$ is coverable in $\mathit{RBN}(\mathcal{P})$, i.e., there exist a reachable graph $\theta$ and a node $v \in \theta$ such that $L(\theta)(v) = s_{n-1}$. Since $LP$ is a path in $\mathcal{P}'_0$, the transition $s_{n-1} \rightarrow s_n$ has to be a broadcast transition labelled by some symbol $!!a$. Hence the node $v$ can broadcast $!!a$ and move into the configuration $s_n = s$.
	
	Suppose $i > 0$. Again since $s$ is reachable in 
	$\mathcal{P}'_i$, there exists a path $LP = s_0 \rightarrow s_1 \rightarrow \cdots \rightarrow s_{n-1} \rightarrow s_n = s$ in $\mathcal{P}'_i$. We prove the claim by a second induction on $n$. For the base case of $n = 0$, it is once again clear that $s_0$ is an initial configuration and so it is coverable in $\mathit{RBN}(\mathcal{P})$. Suppose $n > 0$. By our secondary induction hypothesis, there exists a path in $\mathit{RBN}(\mathcal{P})$ of the form $\theta_0 \rightarrow \theta_1 \rightarrow \cdots \rightarrow \theta_m$ and a node $v \in \theta_m$ such that $L(\theta_m)(v) = s_{n-1}$. We now consider two cases: Suppose $s_{n-1} \rightarrow s_n$ is a broadcast transition labelled by $!!a$. It is then clear that $v$ can broadcast $!!a$ to move into the configuration $s_n = s$. 
	
	Otherwise, $s_{n-1} \rightarrow s_n$ is a receive transition labelled by some letter $??a$. Since this transition belongs to $\mathcal{P}'_i$ it must have been added to the set $\mathtt{AddT}_j$ for some $j \le i$. But notice that we add a new receive transition labelled by $??a$ in the $j^{th}$ iteration iff there exists  $c \in C_a$ such that
	$c$ is coverable in the transition system $\mathcal{P}'_{i-1}$. Therefore, by definition of coverability there exists $s' \ge c$ such that $s'$ is reachable in $\mathcal{P}'_{i-1}$. By our primary induction hypothesis, $s'$ is coverable in $\mathit{RBN}(\mathcal{P})$. So let $\theta'_0 \rightarrow \theta'_1 \rightarrow \cdots \rightarrow \theta'_l$ be a path in $\mathit{RBN}(\mathcal{P})$ and let $v' \in \theta'_l$ be such that $L(\theta'_l)(v') = s'$. Notice that by the property of compatibility, there is a broadcast transition labelled by $!!a$ which is enabled at $s'$.
	
	Now consider the initial graphs $\theta_0$ and $\theta_0'$. Execute the first run from $\theta_0$ so that it reaches the graph configuration $\theta_m$. Now, execute the second run from the initial graph $\theta_0'$ so that it reaches the graph configuration $\theta'_l$. This can be done since these two executions are independent of each other. Now add a link between $v$ and $v'$ and broadcast the message $a$ from $v'$. Hence $v$ will receive the message $a$ and will move into the configuration $s_n = s$. 
\end{proof}

\begin{lemma} \label{lemma2}
	If $s$ is coverable in the reconfigurable broadcast network $\mathit{RBN}(\mathcal{P})$, then $s$ is reachable in $\mathcal{P}'_w$.
\end{lemma}

\begin{proof}
	Suppose $s$ is coverable in the reconfigurable broadcast network $\mathit{RBN}(\mathcal{P})$. Therefore, there exists an initial graph $\theta_0$, a path $LP = \theta_0 \rightarrow \theta_1 
	\rightarrow \cdots \rightarrow \theta_{n-1} \rightarrow \theta_n$ and a node $v$ such that $L(\theta_n)(v) = s$. We will prove the claim by induction on $n$. The claim is clear for the base case of $n = 0$.
	
	Suppose $n > 0$. Let $s'$ be the configuration $L(\theta_{n-1})(v)$. If $s' = s$, then by the induction hypothesis we are done. Suppose $s' \neq s$. Therefore, there should be a transition from $s'$ to $s$. We now have two cases:
	
	\begin{itemize}
		\item Suppose $s' \rightarrow s$ is a broadcast transition labelled by $!!a$. By induction hypothesis, $s'$ is reachable in $\mathcal{P}'_w$. Since all broadcast transitions are present in $\mathcal{P}'_w$, it follows that $s$ is reachable in $\mathcal{P}'_w$ as well.
		
		\item Suppose $s' \rightarrow s$ is a receive transition labelled by $??a$. Hence the node $v$ in $V(\theta_{n-1})$ received a message $a$ and so there should have been a node $u \in V(\theta_{n-1})$ in configuration $s_u'$ such that $u$ broadcasted a message $a$ to reach some configuration
		$s_u$ in the graph $\theta_n$. By induction hypothesis, $s_u'$ is coverable in the transition system $\mathcal{P}'_w$. Hence there exists at least one transition with the broadcast label $!!a$ which is enabled in $\mathcal{P}'_w$. This means that there exists at least one configuration $c \in C_a$ such that $c$ is coverable in $\mathcal{P}'_w$. Hence for the letter $a$, line 10 of the algorithm will eventually become true and so the transition $s' \rightarrow s$ would have been added to $\mathcal{P}'_w$. This means that the transition $s' \rightarrow s$ is present in $\mathcal{P}'_w$. By induction hypothesis, $s'$ is reachable in $\mathcal{P}'_w$ and so $s$ is reachable as well.
	\end{itemize}
\end{proof}

Hence, we have

\begin{theorem}
	Coverability in reconfigurable well-structured broadcast networks is decidable.
\end{theorem}

\begin{proof}
	Notice that the algorithm returns its answer based on whether the given configuration $s$ is coverable in $\mathcal{P}'_w$ or not. 
	
	Suppose $s$ is coverable in $\mathcal{P}'_w$. Therefore, 
	there exists $s' \ge s$ such that $s'$ is reachable in $\mathcal{P}'_w$ and so by Lemma \ref{lemma1}, $s'$ is coverable in $\mathit{RBN}(\mathcal{P})$. Therefore $s$ is coverable in $\mathit{RBN}(\mathcal{P})$. The other side of the proof follows by a similar argument involving Lemma \ref{lemma2}.
\end{proof}

We make a small remark on the complexity of the above algorithm. 
Notice that the main bottleneck in the running time of this algorithm are the coverability tests to the transition system $\mathcal{P}'$. We claim that the coverability problem for reconfigurable broadcast networks of any class of WSTS cannot be faster than the coverability problem for that class. Indeed if such a faster algorithm were to exist we can do the following: Given any labelled WSTS $\mathcal{P}$ in that class, interpret all the labels as broadcast transitions and get a reconfigurable network $\mathit{RBN}(\mathcal{P})$. It is easy to see that a configuration is reachable in $\mathit{RBN}(\mathcal{P})$ iff it is reachable in $\mathcal{P}$.
Now running the faster algorithm on $\mathit{RBN}(\mathcal{P})$ leads to a contradiction. On the other hand it is easy to see that if we can compute the set $C_a$ for each letter quickly, then the number of coverability tests that we ask to $\mathcal{P}'$ is atmost $O(C^2)$ where $C = \sum_{a \in \Sigma} |C_a|$. Hence, the algorithm for coverability in reconfigurable networks runs in time $O(C^2 \times Cov(\mathcal{P}))$ where $Cov(\mathcal{P})$ is the running time of the coverability algorithm for $\mathcal{P}$. A similar argument holds for the amount of space required by the algorithm as well. Since the coverability problem for VASS is \textsc{EXPSPACE}-complete \cite{CovPetriNets}, it follows by the above arguments that

\begin{theorem}
	Coverability of reconfigurable VASS broadcast networks is \textsc{EXPSPACE}-complete.
\end{theorem}

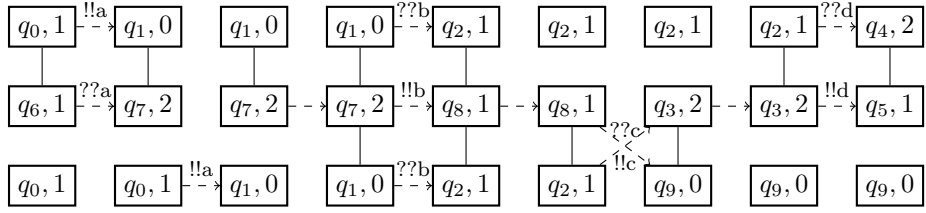
\begin{figure}
	\begin{tikzpicture}
	\begin{scope}[every node/.style={draw,rectangle,thick,minimum size = 4mm},node distance = 0.5cm]
	
		\node(v00) {$q_0,1$};
		\node(v10) [below = of v00] {$q_6,1$};
		\node(v20) [below = of v10] {$q_0,1$};
		\draw (v00) -- (v10);
	
		\node(v01) [right = of v00] {$q_1,0$};
		\node(v11) [right = of v10] {$q_7,2$};
		\node(v21) [right = of v20] {$q_0,1$};
		\draw (v01) -- (v11);
	
		\node(v02) [right = of v01] {$q_1,0$};
		\node(v12) [right = of v11] {$q_7,2$};
		\node(v22) [right = of v21] {$q_1,0$};
		\draw (v02) -- (v12);
	
		\node(v03) [right = of v02] {$q_1,0$};
		\node(v13) [right = of v12] {$q_7,2$};
		\node(v23) [right = of v22] {$q_1,0$};
		\draw (v03) -- (v13);
		\draw (v13) -- (v23);
	
		\node(v04) [right = of v03] {$q_2,1$};
		\node(v14) [right = of v13] {$q_8,1$};
		\node(v24) [right = of v23] {$q_2,1$};
		\draw (v04) -- (v14);
		\draw (v14) -- (v24);
	
		\node(v05) [right = of v04] {$q_2,1$};
		\node(v15) [right = of v14] {$q_8,1$};
		\node(v25) [right = of v24] {$q_2,1$};
		\draw (v15) -- (v25);
	
		\node(v06) [right = of v05] {$q_2,1$};
		\node(v16) [right = of v25] {$q_9,0$};
		\node(v26) [right = of v15] {$q_3,2$};
		\draw (v16) -- (v26);
	
		\node(v07) [right = of v06] {$q_2,1$};
		\node(v17) [right = of v26] {$q_3,2$};
		\node(v27) [right = of v16] {$q_9,0$};
		\draw (v07) -- (v17);
	
		\node(v08) [right = of v07] {$q_4,2$};
		\node(v18) [right = of v17] {$q_5,1$};
		\node(v28) [right = of v27] {$q_9,0$};
		\draw (v08) -- (v18);
	
	\end{scope}
	
		\draw[dashed,->] (v10) -- (v11) node[midway,above] {\footnotesize{??a}};
		\draw[dashed,->] (v00) -- (v01) node[midway,above] {\footnotesize{!!a}};
	
		\draw[dashed,->] (v21) -- (v22) node[midway,above] {\footnotesize{!!a}};
	
		\draw[dashed,->] (v12) -- (v13);
	
		\draw[dashed,->] (v03) -- (v04) node[midway,above] {\footnotesize{??b}};
		\draw[dashed,->] (v13) -- (v14) node[midway,above] {\footnotesize{!!b}};
		\draw[dashed,->] (v23) -- (v24) node[midway,above] {\footnotesize{??b}};
	
		\draw[dashed,->] (v14) -- (v15);
	
		\draw[dashed,->] (v15) -- (v16) node[midway,below] {\footnotesize{!!c}};
		\draw[dashed,->] (v25) -- (v26) node[midway,above] {\footnotesize{??c}};
	
		\draw[dashed,->] (v26) -- (v17);
	
		\draw[dashed,->] (v17) -- (v18) node[midway,above] {\footnotesize{!!d}};
		\draw[dashed,->] (v07) -- (v08) node[midway,above] {\footnotesize{??d}};

	\end{tikzpicture}
	\caption{Covering $q_4$ in $\mathit{RBN}$}
	\label{exec}
\end{figure}

Figure \ref{exec} demonstrates a run in the reconfigurable broadcast network specified by the process in Figure \ref{VASS process}. Recall that reconfigurations are necessary in this case to cover $q_4$.\\

Finally, introducing \textit{arbitrary reconfigurations} in the model might not seem too realistic. But in fact, with respect to \textit{coverability}, this model is equivalent to:

\begin{enumerate}
	\item Static topology with intermittent nodes, i.e., a topology in which there are no reconfigurations but nodes can crash and 
	restart in the \textbf{same} control state in which it crashed. \cite{errors}
	\item Static topology with message loss, i.e., a topology in which there are no reconfigurations but messages may get lost arbitrarily. 
	\cite{errors}
	\item Asynchronous broadcast network with a bag model. \cite{asynch}
	\item Asynchronous broadcast network with a lossy FIFO queue. \cite{asynch}
	\item Globally constrained runs, i.e., a run in which the number of reconfigurations allowed in between two broadcasts can be atmost $k \ge 1$.
	\cite{constraint}
	\item Locally constrained runs, i.e., a run in which the number of reconfigurations \textbf{each node} is allowed to make in between two 
	broadcasts can be atmost $k \ge 1$.
	\cite{constraint}
\end{enumerate}

The proofs given in these papers are for the case when the processes are finite state systems. But these claims can be proved for the infinite state case as well, by noticing that the corresponding proofs go through even in the case of infinite state systems. Intuitively, this is because the equivalence proofs only manipulate the graph topology of the underlying model.

\section{Pushdown broadcast networks}

In the previous section, we gave an algorithm to decide coverability of reconfigurable broadcast networks whenever the underlying process is well-structured. In this section, we will see that a minor modification of the algorithm will also give rise to a coverability algorithm for reconfigurable networks whenever the underlying process is a pushdown automaton. We briefly recall the necessary definitions and theorems for pushdown automata.

For our purposes, a pushdown specification is a tuple $(Q, \Sigma_b \cup \Sigma_r, \Gamma, Q_0, \Delta)$ where $Q$ is a finite set of states, $\Gamma$ is a finite set called the \textit{stack alphabet}, $Q_0$ is a subset of $Q$ called the \textit{initial states} and $\Delta$ is of the form $\Delta \subseteq Q \times \Sigma_b \cup \Sigma_r \times \Gamma \cup \{\epsilon\} \times Q \times \Gamma^{*}$. We assume that $\Gamma$ has a special $\bot$ symbol such that $\bot$ is the end of stack symbol which is neither pushed nor popped by any of the transitions.

This specification describes a \textit{pushdown transition system} $\mathcal{P} = (S, \Sigma_b \cup \Sigma_r, S_0, R)$, where $S = Q \times \Gamma^{*}$ is the set of configurations, $S_0 = Q_0 \times \{\bot\}$ is the set of initial configurations and the transition relation $R \subseteq S \times \Sigma_b \cup \Sigma_r \times S$, where $(s,a,s') \in R$ iff $\exists (q,a,g,q',h) \in \Delta$ such that $s = (q,gw)$ and $s' = (q',hw)$. Given a transition $t \in \Delta$ such that $t = (q,a,g,q',h)$, let $c_t$ denote the configuration $(q,g)$. Let $C_a = \{c_t : t \in \Delta, \ t \text{ is labelled by } !!a\}$

Define an order among the configurations as follows: $s \le s'$ iff $s = (q,w), s'= (q,w')$ and $w$ is a prefix of $w'$. Notice that the prefix relation on words is \textbf{not} a well-quasi ordering.  The coverability (or the control state reachability) problem for pushdown transition systems is the following: Given a configuration $s$, decide if there exists $s_0 \in S_0$ and $s' \ge s$ such that $s_0 \xrightarrow{*} s'$. It is known that the coverability problem is decidable. In fact, the coverability problem is solvable in polynomial time \cite{Pushdown}.

With these definitions of specification and $C_a$ we claim that Algorithm 1 would also solve the coverability problem for reconfigurable pushdown broadcast networks. Indeed, slight modifications of Lemmas \ref{lemma1} and \ref{lemma2} also hold for pushdown transition systems. In the sequel, we just describe the main differences needed to be made to the original proofs to get decidability of coverability for reconfigurable pushdown broadcast networks.

In Lemmas \ref{lemma1} and \ref{lemma2}, the only time we invoke properties of WSTS for the underlying process is for the decidability of coverability and compatibility of transitions. The former condition is not a problem for pushdown processes as coverability is decidable. For compatibility, we have to show that if $s' \ge s$ and there exists a transition $(s,a,c) \in R$ then there also exists a transition $(s',a,c') \in R$ such that $c' \ge c$. By definition of $(s,a,c) \in R$, there exists $(q,a,g,q',h) \in \Delta$ such that $s = (q,gw)$ and $c = (q',hw)$. Since $s' \ge s$, it follows that $s'$ is of the form $s' = (q,gwv)$ where $v \in \Gamma^*$. Hence, the transition $(q,a,g,q',h)$ is also enabled at $s'$ and so $(s',a,c') \in R$ where $c' = (q',hwv) \ge c$. 

We finish with a final discussion on complexity. It is not hard to see that the number of coverability tests that we ask to the underlying pushdown transition system is atmost $O(|\Delta| \times |\Sigma|)$. Since coverability of pushdown systems can be done in polynomial time, it follows that the algorithm for reconfigurable pushdown broadcast networks runs in polynomial time.

Hence we get

\begin{theorem}
	Coverability in reconfigurable pushdown broadcast networks is in \textsc{P}.
\end{theorem}

\section{Discussion and open problems}

We have proved that the coverability problem for reconfigurable broadcast networks is decidable for well-structured processes and pushdown processes. 
Two more classic problems considered in the literature for finite state processes are the \textit{target} and \textit{repeated coverability} problems. We phrase these two problems in the context of well-structured processes. 

Given a well-structured process $\mathcal{P} = (S,\Sigma_b \cup \Sigma_r,S_0,R,\le)$, the \textit{target} problem is the following: Given an upward closed set of configurations $U \subseteq S$, decide if there exists a run in $\theta_0 \rightarrow \theta_1 \rightarrow \cdots \rightarrow \theta_m$ such that $L(\theta_m) \subseteq U$. The \textit{repeated coverability} problem is the following: Given a configuration $s$, decide if there exists an infinite run $\theta_0 \rightarrow \theta_1 \rightarrow \cdots$ and an increasing subsequence 
$i_1 < i_2 < \cdots$ such that for each $\theta_{i_j}$, there exists a configuration $s_{i_j} \ge s$ such that $s_{i_j} \in L(\theta_{i_j})$. 

It is known that for finite state processes the target problem is undecidable in the case of static and bounded path topologies \cite{static}. When reconfigurations are allowed, the problem becomes decidable \cite{mobile}. We think that the target problem becomes undecidable for well-structured processes in the reconfigurable case, however we do not have a proof for the same.

For static topologies, the repeated coverability problem becomes undecidable even for finite state systems. This remains true even when we restrict the set of graphs to be path bounded \cite{static}. However, repeated coverability once again becomes decidable for finite state systems when reconfigurations are allowed \cite{latest}. The same is not the case for well-structured processes. Since repeated coverability for well-structured transition systems is undecidable in general \cite{lcs}, it follows that the repeated coverability problem for reconfigurable well-structured broadcast networks is also undecidable. However, repeated coverability is decidable for a class of WSTS called \textit{very-WSTS} \cite{very-WSTS}. It would be interesting to see if repeated coverability for reconfigurable networks is decidable when the underlying process is a very-WSTS.

\section{Conclusion}

In this paper, we have defined broadcast networks for well-structured processes and proved decidability of coverability for 
various types of semantics. In particular, we have given an algorithm to determine if a given configuration can be covered in 
any run under the reconfiguration semantics. We have also studied decision procedures for various classes of restricted topologies 
which include the set of all path bounded graphs, the set of all cliques, and the set of all graphs with bounded diameter and degree.
A notable ingredient in these decision procedures is the construction of another well-structured transition system to decide 
coverability of configurations.


\bibliographystyle{plainurl}
\bibliography{ref}

\end{document}